\documentclass[fleqn,usenatbib]{mnras}

\usepackage{newtxtext,newtxmath}
\usepackage[T1]{fontenc}
\usepackage{ae,aecompl}
\usepackage{graphicx}	% Including figure files
\usepackage{amsmath}	% Advanced maths commands
\usepackage{svg}
\usepackage{gensymb}
\usepackage{xspace}
\usepackage[export]{adjustbox} % for aligning figures left or right
\usepackage{xcolor}
\usepackage{orcidlink}
\usepackage{mathtools}

\newcommand{\phn}{$\hphantom{0}$}
\newcommand{\phs}{$\hphantom{-}$}
\newcommand{\Gaia}{{\sl Gaia}\xspace}

\newcommand{\perpix}{\mbox{pixel$^{-1}$}}

\newcommand{\masyr}{\hbox{mas\,yr$^{-1}$}}

\newcommand{\Mtot}{\mbox{$M_{\rm tot}$}}
\newcommand{\Lbol}{\mbox{$L_{\rm bol}$}}
\newcommand{\Teff}{\mbox{$T_{\rm eff}$}}
\newcommand{\logg}{\mbox{$\log(g)$}}

\newcommand{\Msun}{\mbox{$M_{\sun}$}}
\newcommand{\Mjup}{\mbox{$M_{\rm Jup}$}}

\newcommand{\Rjup}{\mbox{$R_{\rm Jup}$}}
\newcommand{\Lsun}{\mbox{$L_{\sun}$}}

\newcommand{\vhs}{VHS~J1256$-$1257}
\newcommand{\vhslong}{VHS~J125601.92$-$125723.9}

\title[VHS~J1256$-$1257AB~b Masses, Age \& Orbits]{On the Masses, Age, \& Architecture of the VHS~J1256$-$1257AB~b System}

\author[T. J. Dupuy et al.]{Trent J.~Dupuy$^{1}$\thanks{E-mail: tdupuy@roe.ac.uk}\,\orcidlink{0000-0001-9823-1445},
Michael C.\ Liu$^{2}$\,\orcidlink{0000-0003-2232-7664},
Elise L.\ Evans$^{1}$,
William M.\ J.\ Best$^{3}$\,\orcidlink{0000-0003-0562-1511},
Logan A.\ Pearce$^{4}$\thanks{NSF Graduate Research Fellow}\,\orcidlink{0000-0003-3904-7378}, \newauthor
Aniket Sanghi$^{2, 3}$\,\orcidlink{0000-0002-1838-4757},
Mark W.\ Phillips$^{2}$\,\orcidlink{0000-0001-6041-7092},
and Daniella C.\ Bardalez Gagliuffi$^{5}$\,\orcidlink{0000-0001-8170-7072}
\\
$^{1}$Institute for Astronomy, University of Edinburgh, Royal Observatory, Blackford Hill, Edinburgh, EH9 3HJ, UK\\
$^{2}$Institute for Astronomy, University of Hawaii, 2680 Woodlawn Drive, Honolulu, HI 96822, USA\\
$^{3}$Department of Astronomy, The University of Texas at Austin, 2515 Speedway C1400, Austin, TX 78712, USA\\
$^{4}$Steward Observatory, University of Arizona, Tucson, AZ 85721, USA \\
$^{5}$Department of Physics \& Astronomy, Amherst College, 25 East Drive, Amherst, MA 01003, USA
}
\pubyear{2022}
\begin{document}
\label{firstpage}
\pagerange{\pageref{firstpage}--\pageref{lastpage}}
\maketitle
\begin{abstract}
VHS~J1256$-$1257~AB is an ultracool dwarf binary that hosts a wide-separation planetary-mass companion that is a key target of the {\sl JWST} Exoplanet Early Release Science (ERS) program. Using Keck adaptive optics imaging and aperture masking interferometry, we have determined the host binary's orbit ($a=1.96\pm0.03$\,au, $P=7.31\pm0.02$\,yr, $e=0.883\pm0.003$) and measured its dynamical total mass ($0.141\pm0.008$\,$M_{\odot}$). This total mass is consistent with VHS~J1256$-$1257~AB being a brown dwarf binary or pair of very low-mass stars. In addition, we measured the orbital motion of VHS~J1256$-$1257~b with respect to the barycenter of VHS~J1256$-$1257~AB, finding that the wide companion's orbit is also eccentric ($e=0.68^{+0.11}_{-0.10}$), with a mutual inclination of $115\degree\pm14\degree$ with respect to the central binary. This orbital architecture is consistent with VHS~J1256$-$1257~b attaining a significant mutual inclination through dynamical scattering and thereafter driving Kozai-Lidov cycles to pump the eccentricity of VHS~J1256$-$1257~AB. We derive a cooling age of $140\pm20$\,Myr for VHS~J1256$-$1257~AB from low-mass stellar/substellar evolutionary models. At this age, the luminosity of VHS~J1256$-$1257~b is consistent with both deuterium-inert and deuterium-fusing evolutionary tracks. We thus find a bimodal probability distribution for the mass of VHS~J1256$-$1257~b, either $12.0\pm0.1$\,$M_{\rm Jup}$ or $16\pm1$\,$M_{\rm Jup}$, from these models. Future spectroscopic data to measure isotopologues such as HDO and CH$_3$D could break this degeneracy and provide a strong test of substellar models at the deuterium-fusion mass boundary.
\end{abstract}
\begin{keywords}
astrometry -- binaries: visual -- brown dwarfs -- exoplanets -- planetary systems
\end{keywords}

\section{Introduction}

Two fundamental parameters govern the bulk properties of gas-giant planets and brown dwarfs: mass and age. Mass is difficult to measure directly for imaged planets because of their long orbital periods, though there has been progress for a few planets inside 20\,au \citep{2018NatAs.tmp..114S, 2019ApJ...871L...4D, 2022MNRAS.509.4411D, Brandt_2021_beta_Pic_bc, 2021ApJ...915L..16B}. Ages for imaged planets have been largely reliant on an object belonging to a well-studied young association because of the limitations of determining precise ages for field stars. One notable exception is the Y-dwarf WD~0806-661~b \citep{2011ApJ...730L...9L} whose age $1.5^{+0.5}_{-0.3}$\,Gyr is determined by the cooling time of its white-dwarf host.

There are a handful of gas-giant companions with the potential for precise age dating using cooling ages from their low-mass hosts. Such ages are similar to white-dwarf cooling ages but without the need to estimate a stellar progenitor's lifetime. Because both stars and brown dwarfs begin with an initial entropy that is related to their mass, the age of such a low-mass object can be determined by measuring its mass and present-day luminosity \citep[e.g.,][]{2008ApJ...689..436L,2009IAUS..258..317B}. Gas giants in systems where the host stars are binary brown dwarfs or pre-main--sequence stars are amenable to such cooling-age measurements.  \vhslong\ (hereinafter \vhs~AB) is one of the few such host binaries.

\citet{2015ApJ...804...96G} used the Visible and Infrared Survey Telescope for Astronomy (VISTA) Hemisphere Survey (VHS) to discover that VHS~J125601.58$-$125730.3 (hereinafter \vhs~b) is a common-proper motion companion to \vhs~AB, at a projected separation of $8\farcs06$. They derived spectral types and gravity classifications for the host and companion of M$7.5\pm0.5$~\textsc{int-g} and L$7.0\pm1.5$~\textsc{vl-g}, respectively. They measured a parallax of $78.8\pm6.4$\,mas, which placed the companion in the same location as HR~8799~b on the color-magnitude diagram. The parallax has since been updated, first by the Hawai`i Infrared Parallax Program \citep[$45.0\pm2.4$\,mas;][]{2020RNAAS...4...54D} and most recently by Gaia~EDR3 \citep[$47.27\pm0.47$\,mas = $21.14\pm0.22$\,pc;][]{2016A&A...595A...1G,2021A&A...649A...1G}, making the system more distant than originally thought. The companion is no longer a direct analog to HR~8799~b, but given the primary's age \citep[150--300~Myr;][]{2020RNAAS...4...54D}, it is still potentially planetary mass. Its cool temperature and wide separation make \vhs~b appealing for direct imaging studies, including being the primary spectroscopy target for {\sl JWST}'s Exoplanet Early Release Science program \citep{2022PASP..134i5003H}.

Adaptive optics (AO) imaging revealed that the host is a binary \citep{2016ApJ...818L..12S, 2016ApJ...830..114R}, making \vhs\ a rare triple system potentially composed entirely of substellar objects.  The inner binary's projected separation at discovery was $2.62\pm0.03$\,au (using the latest distance), which would correspond to a $\sim$10-year orbital period. The likelihood of obtaining dynamical masses on such a relatively short time scale motivated us to begin an orbit monitoring campaign, as we have done for other substellar binaries \citep[e.g.,][]{2017ApJS..231...15D}. We present here the dynamical masses for the binary components and a corresponding cooling age that suggests the directly imaged companion \vhs~b may be below the deuterium-fusing mass limit.

\begin{table}
\centering
\caption[]{Keck/NIRC2 Relative Astrometry of \vhs~AB.} \label{tbl:relast}
\begin{tabular}{lccrr}
\hline
Epoch & Sep (mas) & PA (\degree) & Corr & $\Delta{K_{\rm MKO}}$ (mag) \\
\hline
2016.059  &     $128.21\pm 0.14$ & $168.07\pm0.05$ & $ 0.01$ & $ 0.021 \pm0.006$     \\
2017.050  &     $139.7 \pm 2.8 $ & $163.5 \pm1.9 $ & $-0.20$ & $-0.06  \pm0.16 $\phn \\
2017.220  &     $140.5 \pm 0.7 $ & $162.9 \pm0.4 $ & $ 0.35$ & $ 0.031 \pm0.023$     \\
2018.017  &     $135.84\pm 0.29$ & $158.37\pm0.15$ & $ 0.83$ & $ 0.026 \pm0.026$     \\
2019.259  &     $109.6 \pm 0.4 $ & $150.36\pm0.11$ & $ 0.32$ & $ 0.032 \pm0.028$     \\
2021.018  & \phn$ 35.02\pm 0.26$ & $112.5 \pm0.6 $ & $ 0.00$ & $ 0.033 \pm0.005$     \\
2022.066  & \phn$ 70.0 \pm 1.8 $ & $181.0 \pm1.7 $ & $ 0.64$ & $ 0.19  \pm0.04 $\phn \\
2022.271  & \phn$ 85.6 \pm 0.7 $ & $177.2 \pm0.4 $ & $-0.52$ & $ 0.25  \pm0.05 $\phn \\
\hline
\end{tabular}
\end{table}

\section{Observations} \label{sec:obs}

We obtained astrometry for the \vhs\ system from the Keck~II Telescope using the facility AO system. We began monitoring on 2016~Jan~22~UT using the Maunakea Observatories $K$-band filter \citep{2002PASP..114..180T} and NIRC2's narrow-camera mode, which has a pixel scale of $9.971\pm0.004$\,mas\,\perpix \citep{2016PASP..128i5004S} and field-of-view of $10\farcm2\times10\farcm2$. Most of our measurements were made using the standard laser guide star (LGS) AO system \citep{2006PASP..118..297W}, which uses a Shack-Hartmann wavefront sensor to measure the LGS and a separate red-optical sensor observing \vhs~AB itself as a tip-tilt reference. At one epoch, 2022~Jan~24~UT, we instead used the infrared pyramid wavefront sensor \citep{2020JATIS...6c9003B}, with \vhs~AB providing  natural guide star AO correction.

Our first imaging from 2016 was obtained less than a year after the discovery imaging using MagAO/Clio2 and NIRC2 from \citet{2016ApJ...818L..12S}, and it was consistent with their measurement of increasing projected separation. In the following, we use the earliest data that comes from MagAO along with our own NIRC2 data. By 2018 the projected separation began decreasing, and eventually, the binary was unresolved on 2021~Jan~5~UT. We obtained aperture masking interferometry data the following night, using the 9-hole mask \citep{Ireland:2008yq}, and successfully resolved \vhs~AB at 35\,mas separation. In addition, starting with our first observation in 2016, we regularly obtained our imaging in a way that captured both \vhs~AB as well as \vhs~b. This allowed us to measure astrometry for \vhs~AB and the orbital motion of \vhs~b.

Our methodology for reducing NIRC2 imaging and masking data is described extensively in our previous work \citep[e.g.,][]{2017ApJS..231...15D}. Briefly, we perform standard calibrations (dark subtraction and flat-fielding with dome flats) and then measure the separation, position angle (PA), and flux ratio in individual images. This is done using StarFinder \citep{2000A&AS..147..335D} when possible, but when this fails at the closest separations we use an analytical, multi-component Gaussian PSF model optimized using the Levenberg-Marquardt algorithm implemented in IDL by the \textsc{mpfit} routine \citep{2009ASPC..411..251M}. We also tested a Moffat PSF model because \citet{2012CardosoC} and \citet{2022AJ....163..288C} showed it is the optimal profile for NACO data, but it did not significantly alter our results. We correct our measured pixel positions for NIRC2's distortion using the \citet{2016PASP..128i5004S} astrometric calibration, which also provides the pixel scale and orientation of the images. Final measurements at an epoch are the mean and standard deviation of values from individual images. For masking data, we obtain binary parameters by fitting the closure phases using the Sydney pipeline \citep{Ireland:2008yq}.\footnote{\url{https://github.com/mikeireland/idlnrm}} 

Table~\ref{tbl:relast} presents our relative astrometry for \vhs~AB, including the linear Pearson correlation coefficient (Corr) for separation and PA. Table~\ref{tbl:absast} presents astrometry of \vhs~b we derived from imaging epochs where \vhs~AB was sufficiently well resolved for StarFinder analysis and that conformed to a standard configuration with the NIRC2 $y$-axis ${\rm PA}\approx0$\degree\ and \vhs~AB at NIRC2 $(x,y) \approx (250,800)$\,pix. By keeping all three components in approximately the same location on NIRC2, the astrometry should be minimally impacted by the $\approx$1\,mas uncertainty in the distortion solution. We follow convention in referring to relative declination as $\Delta\delta$ and right ascension as $\Delta\alpha^* \equiv \Delta\alpha\cos\delta$.

\begin{table}
\centering
\caption[]{Keck/NIRC2 Relative Astrometry of \vhs~b.} \label{tbl:absast}
\begin{tabular}{lcccc}
\hline
Epoch & $\Delta{\alpha^*}_{\rm b-A}$ (mas) & $\Delta{\delta}_{\rm b-A}$ (mas) \\
\hline
2016.059  &  $-4974.0\pm2.8$ & $-6409.8\pm2.3$ \\
2017.220  &  $-4977.6\pm2.2$ & $-6412.8\pm2.1$ \\
2018.017  &  $-4982.9\pm2.5$ & $-6406.5\pm2.6$ \\
2022.271  &  $-5049.0\pm1.3$ & $-6387.5\pm1.1$ \\
\hline
\end{tabular}
\begin{list}{}{}
\item[Note.]-- Astrometry relative to \vhs~A not the barycenter of AB. As described in Section~\ref{sec:orbit}, we determine the position of \vhs~b relative to the \vhs~AB barycenter to be $\Delta(\alpha^*,\delta)_{\rm b-AB} = (-5025.6\pm2.0,-6350\pm9)$\,mas at the mean epoch 2019.853.
\end{list}
\end{table}

Our multi-epoch NIRC2 data precisely constrain the $K$-band flux ratio of \vhs~AB. We used only StarFinder and masking results, as these should be less prone to systematic errors (epochs 2017.05, 2018.02, 2019.26, and 2021.02). The flux ratios are in excellent agreement, with $\chi^2=0.44$ and 3 degrees of freedom (dof), so we adopt the weighted average $\Delta{K_{\rm MKO}} = 0.033\pm0.004$\,mag.

The integrated-light spectrum of \vhs~AB was obtained with IRTF/SpeX in prism mode on 2016~Feb~19~UT as part of NASA IRTF program 2016A079 (PI: Bardalez Gagliuffi). The target was observed at airmass 1.19 with the 0$\farcs$5 slit and $6\times60$\,s exposures. The A0 star HD~112304 was observed immediately after the target for flux calibration and telluric correction. Internal flat fields and argon arc lamps followed the standard observations for wavelength calibration. All data were reduced with the IDL package SpeXtool v4.1. Further details on the observations and instrument settings can be found in \citet{2014ApJ...794..143B} and \citet{2010ApJ...710.1142B}.

\section{Orbit analysis} \label{sec:orbit}

For the three-body system of \vhs~AB~b, we separate our orbital analysis into the relative orbit of the host binary ($\sim$2\,au) and the orbital motion of the wide companion ($\sim$170\,au) relative to the AB barycenter. Dynamical interactions are negligible for such a wide, low mass-ratio ($M_{\rm comp}/M_{\rm host} < 0.1$) system.

To fit the relative orbit of \vhs~AB we used {\sc orvara} \citep[v1.0.4;][]{2021AJ....162..186B}. {\sc orvara} utilizes a novel, highly-efficient eccentric anomaly solver and determines posteriors of orbital parameters using the affine-invariant \citep{2010CAMCS...5...65G} Markov-Chain Monte Carlo (MCMC) sampler {\sc emcee} \citep{2013PASP..125..306F} with parallel-tempering \citep{Vousden_2016_PT}. We provide our {\sc orvara} configuration files as supplementary data here, but briefly, we fitted all eight standard parameters for a relative astrometric fit with their default priors (linear-flat in eccentricity $e$ and viewing angles, except inclination $p(i)\propto\sin{i}$, and log-flat in mass and semimajor axis $a$). Relative astrometry only constrains the total mass, $\Mtot \equiv M_{\rm A}+M_{\rm B}$, so we placed no limiting priors on the component masses. Thus, $M_{\rm A}$ and $M_{\rm B}$ varied freely in the {\sc orvara} MCMC analysis, but were always constrained implicitly to follow a consistent \Mtot. Our results are based on a run with 100 walkers and $10^5$ steps for the MCMC and 5 temperatures for parallel tempering. We thinned our chains, retaining every 50th step, and discarded the first 50\% as burn-in, yielding $10^5$ final samples in our posterior.

\begin{figure*}
 \begin{center}
  \includegraphics[width=\textwidth]{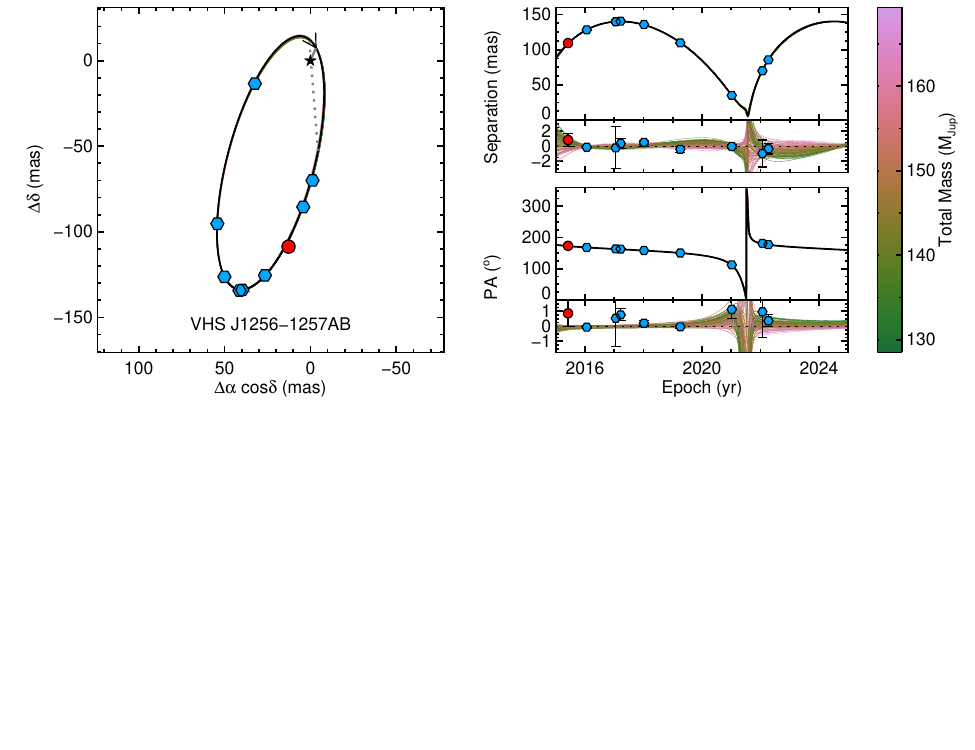}
  \vskip -2.5in
  \caption{Relative astrometry of \vhs~AB from our Keck/NIRC2 data (blue hexagons; Table~\ref{tbl:relast}) and MagAO/Clio2 discovery imaging (red circles). {\it Left:} Positions of \vhs~B (colored symbols; error bars are smaller than the symbols) relative to \vhs~A (black star). The maximum-likelihood orbit (thick black line) is shown with its line of nodes (dotted grey line) and time of periastron passage (solid grey line and arrow) indicated. Thin lines color-coded by the total mass show 50 randomly drawn orbits drawn from our MCMC posterior. {\it Right:} The same measurements and orbits shown as separation (top) and PA (bottom) as a function of time with subpanels showing the residuals with respect to the maximum-likelihood orbit.  \label{fig:skyplot}}
 \end{center}
\end{figure*}

\begin{table}
\setlength{\tabcolsep}{3pt}
\centering
\caption[]{\vhs~AB orbital parameters derived from {\sc orvara}.} \label{tbl:orbit}
\begin{tabular}{lcc}
\hline
Property & Median $\pm$1$\sigma$ & 95.4\% c.i. \\
\hline
Total mass, \Mtot\ (\Msun)                                     & $0.141\pm0.008$             &   0.125, 0.157   \\[3pt]
Semimajor axis, $a$ (au)                                       & $1.96\pm0.03$               &    1.89, 2.03    \\[3pt]
Eccentricity, $e$                                              & $0.8826_{-0.0024}^{+0.0025}$&  0.8776, 0.8875  \\[3pt]
Inclination, $i$ (\degree)                                     & $118.7\pm1.0$\phn\phn       &   116.7, 120.7   \\[3pt]
PA of ascending node, $\Omega$ (\degree)                       & $4.4\pm0.5$                 &     3.5, 5.3     \\[3pt]
Argument of periastron, $\omega$ (\degree)                     & $44.9\pm1.0$\phn            &    42.8, 46.9    \\[3pt]
Mean longitude at $t_{\rm ref}$, $\lambda_{\rm ref}$ (\degree) & $-163.5\pm1.6$\phs\phn\phn  &$-$166.7, $-$160.1\\
\hline														 
Period, $P$ (yr)                                               & $7.307_{-0.024}^{+0.023}$   &   7.262, 7.357   \\[3pt]
Time of periastron, $t_p$ (yr)                                 & $2021.537_{-0.014}^{+0.015}$\phn\phn\phn&2021.507, 2021.566\\[3pt]
$\tau \equiv (t_p-t_{\rm ref})/P$                              & $0.579\pm0.007$             &   0.565, 0.592   \\
\hline
\end{tabular}
\begin{list}{}{}
\item[*] Reference epoch $t_{\rm ref} = 2010.0$ (55197\,MJD).
\item[Note.] Free parameters in the MCMC are shown in the top section. These were used to compute the parameters in the bottom section. All posterior distributions are nearly Gaussian.
\end{list}
\end{table}

Our measured total mass of $0.141\pm0.008$\,\Msun\ suggests that the components of \vhs~AB are possibly brown dwarfs with masses of $74\pm4$\,\Mjup\ if their mass ratio is near unity. Their orbital eccentricity of $0.8826^{+0.0025}_{-0.0024}$ is the highest ever measured for a very low-mass binary \citep[e.g., see][]{2011ApJ...733..122D,2017ApJS..231...15D}.

Independent of the orbit analysis of \vhs~AB we measured the orbital motion of \vhs~b relative to its host's barycenter (denoted b--AB). As mentioned in Section~\ref{sec:obs}, at some epochs we obtained imaging of all three objects in individual NIRC2 images. The relative positions of A and B are typically measured $\sim$10$\times$ more precisely than A or B to the companion b. This allowed us to approximate the errors in \vhs~AB relative astrometry to be negligible compared to those of \vhs~b. Under this assumption, the position of \vhs~b relative to \vhs~A can be written as
\begin{equation}
\hfill
\Delta{\alpha^*}_{\rm b-A} = \Delta{\alpha^*}_{\rm b-AB} + (\mu_{\alpha^*, {\rm b-AB}} \times t) - [(M_{\rm A}/\Mtot) \times \Delta{\alpha^*}_{\rm A-B}]
\hfill
\end{equation} 
\vskip -0.25in
\begin{equation}
~\Delta{\delta}_{\rm b-A} = \Delta{\delta}_{\rm b-AB} + (\mu_{\delta, {\rm b-AB}} \times t) - [(M_{\rm A}/\Mtot) \times \Delta{\delta}_{\rm A-B}],
\end{equation}
where the left-hand side corresponds to the measurements in Table~\ref{tbl:absast}, the $\Delta_{\rm A-B}$ values on the far right side can be derived from Table~\ref{tbl:relast}, and the rest are free parameters.

We used {\sc mpfit} to find the best-fit solution and then used a Monte Carlo approach to derive the uncertainties in the fit by randomly drawing simulated measurements from the best-fit model with scatter equal to the individual input measurements. Unfortunately, the mass ratio is poorly constrained in this analysis ($M_{\rm A}/\Mtot = 0.45\pm0.08$), likely due to the eccentric orbit and the fact that the measurements used here (i.e., when all three components are well resolved) happen to come from a similar phase of the orbit. In contrast, the orbital motion of the companion relative to the \vhs~AB barycenter is well detected at $\mu_{\alpha^*, {\rm b-AB}} = -10.7\pm0.6$\,\masyr\ and $\mu_{\delta, {\rm b-AB}} = 0.6\pm0.7$\,\masyr.

To fit the orbit of \vhs~b relative to the AB barycenter we used the python package {\sc lofti\_gaia} \citep{2020ApJ...894..115P}.\footnote{\url{https://github.com/logan-pearce/lofti_gaia}} {\sc lofti\_gaia} is based on the Orbits-For-The-Impatient \citep[OFTI; ][]{2017AJ....153..229B} rejection-sampling method and fits orbital parameters of resolved binaries in \Gaia using their proper motions and radial velocities if available. Here we adopted the architecture of {\sc lofti\_gaia} to use our measured proper motion of b relative to the AB barycenter at the mean observation epoch 2019.853, rather than \Gaia astrometry at the mean \Gaia epoch. We fitted six orbital parameters: semimajor axis ($a_{\rm b}$), eccentricity ($e_{\rm b}$), inclination ($i_{\rm b}$), argument of periastron ($\omega_{\rm b}$), longitude of ascending node ($\Omega_{\rm b}$), and time of periastron passage ($t_{\rm p, b}$). Total system mass and distance were drawn from normal distributions of $0.152\pm0.010$\,\Msun\ and $21.14\pm0.22$~pc. This system mass is based on our measured mass for \vhs~AB and an estimated mass of $0.011\pm0.006$\,\Msun\ for \vhs~b from our evolutionary model analysis in Section~\ref{sec:evol}. OFTI rejection sampling generates trial orbits by drawing random values for four orbital parameters from priors in $e_{\rm b}$: Uniform on [0,1); $\cos(i_{\rm b})$: Uniform on [-1,1]; $\omega_{\rm b}$: Uniform on [0,2$\pi$]; orbit phase, $(t_{\rm p, b}-2019.853)/P_{\rm b}$: Uniform on [0,1]. OFTI then scales the semimajor axis and rotates $\Omega_{\rm b}$ to match the input data and determines whether to accept or reject a trial by comparing its proper motion in RA and Dec to our measured values. There is no prior on $a_{\rm b}$ or $\Omega_{\rm b}$. 

\begin{figure}
 \begin{center}
  \includegraphics[width=0.49\textwidth]{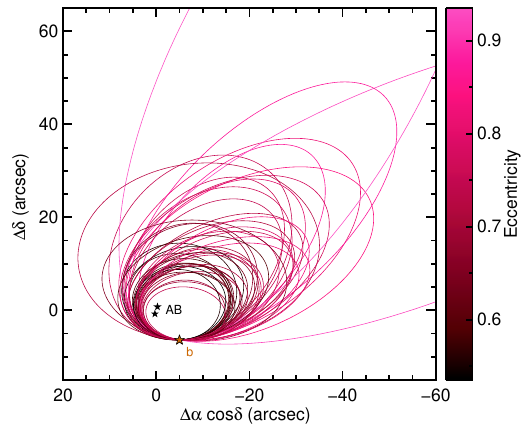}
  \vskip -4pt
  \caption{The orbit of \vhs~b relative to its host binary (black stars; separation of AB relative to b is {\em not to scale}). Shown are 50 randomly drawn orbits from our {\sc lofti\_gaia} analysis, color-coded by eccentricity.
  \label{fig:AB-b-orbit}}
 \end{center}
\end{figure}

We ran {\sc lofti\_gaia} on our measured proper motions until $10^5$ trial orbits were accepted. Table~\ref{tbl:orbit-comp} reports the output probability distributions of orbital parameters of \vhs~b around its host, and Figure~\ref{fig:AB-b-orbit} shows these orbits on the sky.

\begin{table}
\centering
\caption[]{\vhs~b orbital parameters derived from {\sc lofti\_gaia}.} \label{tbl:orbit-comp}
\begin{tabular}{lcc}
\hline
Property & Median $\pm$1$\sigma$ & 95.4\% c.i. \\
\hline
Semimajor axis $a_{\rm b}$ (au)                                       & $350_{-150}^{+110}$       &  \phn150, 1020   \\[3 pt]
Eccentricity $e_{\rm b}$                                              & $0.68_{-0.10}^{+0.11}$    &     0.49, 0.91   \\[3 pt]
Inclination $i_{\rm b}$ (\degree)                                     & $24_{-15}^{+10}$          &    \phn3, 48     \\[3 pt]
Argument of periastron $\omega_{\rm b}$ (\degree)                     & $180_{-130}^{+100}$       &\phn\phn0, 330    \\[3 pt]
PA of the ascending node $\Omega_{\rm b}$ (\degree)                   & $40_{-160}^{+50}$         &   $-$140, 190\phs\\[3 pt]
Time of periastron $t_p, {\rm b}$ (yr)                                & $1240\pm90$\phn\phn       &  \phn980, 1480   \\[3 pt]
\hline
Period $P_{\rm b}$ (kyr)                                              & $16_{-10}^{+7}$           &    \phn4, 82     \\[3 pt]
$\tau_{\rm b} \equiv (t_p-2019.853)/P$                                & $0.047_{-0.039}^{+0.022}$ &    0.000, 0.117  \\[3 pt]
\hline
\end{tabular}
\begin{list}{}{}
\item[Note.] Our {\sc lofti\_gaia} analysis adopted a system mass of $M_{\rm A}+M_{\rm B}+M_{\rm b}=0.152\pm0.010$\,\Msun. Free parameters from the fit are shown in the top section. These were used to compute the parameters in the bottom section.
\end{list}
\end{table}

\section{Luminosities} \label{sec:lbol}

We computed the combined-light bolometric luminosity of \vhs~AB by direct integration of its unresolved optical to mid-infrared (MIR) spectral energy distribution (SED). Our assembled SED consists of available Pan-STARRS-1 \citep[PS1;][]{2016arXiv161205560C} optical photometry ($g$, $r$, $y$), the near-infrared (NIR) IRTF/SpeX prism spectrum from Section~\ref{sec:obs}, NIR photometry from 2MASS \citep{2003tmc..book.....C}, and MIR photometry from the CatWISE catalog \citep[$W1$ and $W2$ bands;][]{2020ApJS..247...69E,2021ApJS..253....8M} and AllWISE catalog \citep[$W3$ and $W4$ bands;][]{2013wise.rept....1C}. We began by flux-calibrating the SpeX spectrum using the weighted average of calibrations derived from PS1 $y$ and 2MASS $JHK_s$ photometry, assuming a systematic noise floor of 0.01~mag for all the filters. We then integrated the flux-calibrated SpeX spectrum to determine the NIR contribution to the bolometric flux, with an error that accounts for the uncertainties in the spectral data points and the overall flux calibration. We determined the optical and MIR contributions to the bolometric flux by simultaneously fitting BT-Settl model atmospheres \citep[CIFIST2011/2015;][]{2012RSPTA.370.2765A, 2015A&A...577A..42B} to the PS1 and WISE photometry (computing synthetic photometry from the models) and the SpeX spectrum (with the models degraded to the non-linear spectral resolution of the 0$\farcs$5 slit). We found the best-fitting BT-Settl model had $\Teff = 2700$\,K and $\logg = 5.0$\,dex. Our final bolometric flux was found by adding the NIR contribution to the integration of the model outside the wavelength range of the SpeX spectrum. The uncertainty in the optical+MIR contribution was obtained from the standard deviation of the corresponding measurements derived using the four model spectra adjacent in \Teff\ and \logg\ to the best-fitting model. Our final bolometric flux of \vhs~AB is $1.47\pm0.04\times10^{-13}$\,W\,m$^{-2}$. Using its parallactic distance of $21.14 \pm 0.22$\,pc, we calculated a bolometric luminosity $\log((\Lbol_{\rm, A}+\Lbol_{\rm, B})/\Lsun) = -2.687\pm0.021$\,dex.

To derive component luminosities for \vhs~AB, we used the empirical relation between $K_s$-band absolute magnitude and \Lbol\ from \citet{2017ApJS..231...15D}. We assumed that $\Delta{K_{\rm MKO}} = \Delta{K_{\rm 2MASS}}$ here because of the near-unity flux ratio. Using a Monte Carlo method, we drew random absolute magnitudes representative of \vhs~A ($9.1\pm0.2$\,mag, truncated at 8.7\,mag, the upper limit of the empirical relation). We then simulated \vhs~B by adding $0.033\pm0.004$\,mag to this absolute magnitude and computed the difference in derived \Lbol\ values from the relation. We found $\log(\Lbol_{\rm,B}/\Lbol_{\rm,A}) = -0.012\pm0.002$\,dex. We therefore calculated component luminosities of $\log(\Lbol_{\rm,A}/\Lsun) = -2.982\pm0.021$\,dex and $\log(\Lbol_{\rm,B}/\Lsun) = -2.994\pm0.021$\,dex.

For \vhs~b, we used the value of $\log(\Lbol_{\rm, b}/\Lsun) = -4.568\pm0.009$\,dex derived by \citet{2023ApJ...946L...6M} integrating over the whole 1--20\,\micron\ spectrum observed by {\sl JWST}, where gaps were covered by with BT-Settl models, and the Gaia~EDR3 parallactic distance was used.

\section{Evolutionary model analysis} \label{sec:evol}

Substellar objects with well-determined luminosities enable precise evolutionary model-derived cooling ages (when mass is known) and masses (when age is known). Some key aspects of evolutionary models are quite uncertain, such as the treatment of clouds. The relatively sparse tests of models with objects of known mass, age, and luminosity have found a mixed bag of agreement and potential problems \citep[e.g.,][]{2014ApJ...790..133D,2018AJ....156..168B,Brandt2021_Six_Masses}, so we note that any mass or age derived from evolutionary models should be treated with corresponding uncertainty. In the following, we use our dynamical mass measurement of \vhs~AB to determine a substellar cooling age for the system and then use this cooling age to estimate the mass of \vhs~b.  

For \vhs~AB, the most appropriate evolutionary models are from \citet{2015A&A...577A..42B}. As in our previous work \citep[e.g.,][]{2017ApJS..231...15D}, we used a Monte Carlo rejection-sampling approach to derive an age probability distribution from input luminosity and mass prior distributions. We assumed a linear-flat prior in age and a log-flat prior in $M_{\rm A}$, drawing random, uniformly distributed values, while simultaneously drawing random values of \Mtot\ from our MCMC posterior. We calculated $M_{\rm B}$ as the difference between \Mtot\ and $M_{\rm A}$. For each age-mass pair, for each component, we computed a model luminosity from bilinear interpolation of the model grid. The probability of any sample being accepted was $p=e^{-(\chi^2-\chi_{\rm min}^2)/2}$, where $\chi^2$ was computed as the sum of comparing our measured luminosities to the model-calculated ones, and $\chi^2_{\rm min}$ was the lowest value among the ensemble of trial values. A sample was accepted if a randomly drawn number $0<u<1$ for a given trial satisfied $p>u$. We then computed other model-derived properties, such as \Teff, using the accepted mass and age samples.

We found a cooling age of $140\pm20$\,Myr for \vhs~AB, with an approximately Gaussian probability distribution. This age is consistent with the nondetection of lithium in its spectrum \citep{2015ApJ...804...96G}, as extreme lithium depletion ($>10^{-4}$) corresponds to the older part of the age posterior at $<2$$\sigma$ according to the \citet{2015A&A...577A..42B} models.

\begin{figure*}
 \begin{center}
  \includegraphics[width=0.49\textwidth]{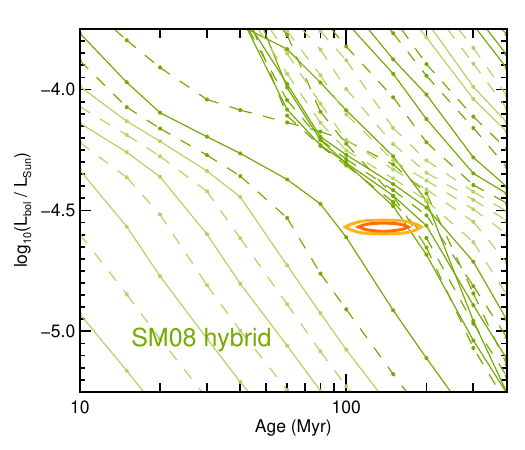}
  \includegraphics[width=0.49\textwidth]{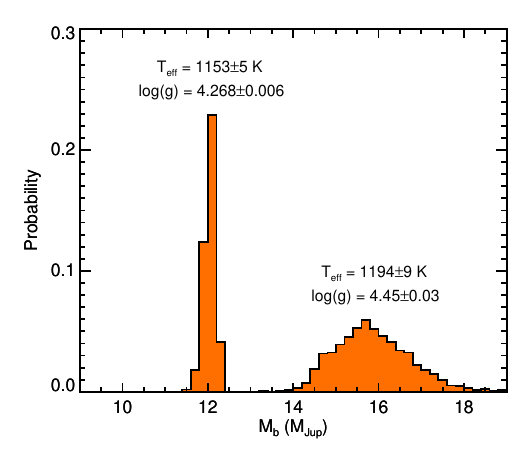}
  \vskip -6pt
  \caption{{\it Left:} evolutionary models from \citet{2008ApJ...689.1327S} predict \Lbol\ as a function of age for objects of varying masses. (Light and dark colored lines alternate between intervals of 0.01\,\Msun, and solid and dashed lines alternate between even and odd masses.) Orange contours indicate the 1$\sigma$ and 2$\sigma$ joint constraints on luminosity and age, lying to the right of the 0.011\,\Msun\ (11.5\,\Mjup) D-inert track and to the left of more massive D-fusing tracks ($>0.014$\,\Msun). (Objects with masses 0.012--0.013\,\Msun\ are predicted to be much more luminous at this age due to a D-fusion spike.) {\it Right:} posterior distribution of the mass of \vhs~b from our rejection sampling Monte Carlo analysis of these models. The distribution is bimodal between a D-inert mass of $12.0\pm0.1$\,\Mjup\ and a D-fusing mass of $16\pm1$\,\Mjup. The lower mass corresponds to a younger age and thus a larger radius, lower \Teff, and lower \logg. \label{fig:evol}}
 \end{center}
\end{figure*}

We used \vhs~AB's age posterior to perform a rejection-sampling analysis of the companion's properties. The only evolutionary model grid that reaches \vhs~b's luminosity and accounts for cloud evolution is the ``hybrid'' grid of \citet{2008ApJ...689.1327S}. As seen in Figure~\ref{fig:evol}, \citet{2008ApJ...689.1327S} models predict that objects with the luminosity and age of \vhs~b should be rare because they fall in a gap between low-mass objects that cannot fuse deuterium and more massive objects that are either fusing deuterium now (and thus more luminous at this age) or have already fused their deuterium (and thus are older at this luminosity). Our rejection sampling analysis correspondingly results in a bimodal posterior distribution. The slightly less probable outcome, with 40\% of the posterior, is that \vhs~b is a deuterium-bearing object of $12.0\pm0.1$\,\Mjup. The slightly more probable outcome is that \vhs~b has already depleted its deuterium and is an object of $16\pm1$\,\Mjup.

The radius of \vhs~b, according to these models, is 1.30\,\Rjup\ in the lower-mass scenario and 1.22\,\Rjup\ in the higher-mass scenario. This translates into slightly different effective temperatures of $1153\pm5$\,K and $1194\pm9$\,K, respectively, as well as surface gravities of $4.268\pm0.006$\,dex and $4.45\pm0.03$\,dex.

The bimodality in \vhs~b's mass drives the properties we derived, so alternative model assumptions could potentially shift the balance significantly in favor of one or the other possibilities. The \citet{2008ApJ...689.1327S} models, in particular, have such a wide gap in the \Lbol--age diagram because the onset of the L/T transition, which these models assume begins at $\Teff=1400$\,K, happens to occur at nearly the same age as deuterium fusion for objects near the deuterium-fusion mass boundary. The L/T transition slows cooling, so objects stay luminous both because of cloud disappearance and deuterium fusion. The L/T transition probably occurs at lower temperatures for low-gravity objects like \vhs~b \citep[e.g.,][]{2006ApJ...651.1166M,2009ApJ...699..168D,2015ApJ...810..158F}, which may significantly impact the size and shape of the deuterium-fusing gap in \Lbol-age space. Naively, such a delayed and lower-\Teff\ L/T transition might be expected to make even lower-mass isochrones have higher luminosities in Figure~\ref{fig:evol}, which would in turn make it more likely that \vhs~b is indeed below the deuterium-fusion mass boundary.

\section{Orbital architecture \& Origins} \label{sec:arch}

We have astrometrically determined the three-dimensional orbits of both the inner host binary (A--B) and the outer companion about its barycenter (AB--b). This allows us to constrain the orbital architecture of the system and thus, potentially, shed light on its origin. One crucial measurement that our orbit determinations enables is the true mutual inclination of the A--B and AB--b orbital planes, 
\begin{equation}
    \hfill
    \cos{i_{\rm AB-b}} = \cos{i_{\rm AB}}\cos{i_{\rm b}} + \sin{i_{\rm AB}}\sin{i_{\rm b}}\cos(\Omega_{\rm AB}-\Omega_{\rm b}).
    \hfill
    \label{eq:cosphi}
\end{equation}
For more detail on how to derive this mutual inclination angle, we point the reader to \citet{2011ApJ...743...61S}. Propagating all measurement uncertainties from {\sc orvara} and {\sc lofti} analyses, we find $i_{\rm AB-b} = 115\pm14$\degree. This reveals that the angular momentum vectors of the two orbital planes are misaligned (8$\sigma$) and also possibly pointing in opposite directions (1.8$\sigma$).

\vhs~AB's orbit is highly eccentric, and one possible explanation for this is that it has been pumped up by Kozai-Lidov cycles \citep{1962AJ.....67..591K,1962P&SS....9..719L}. The observed mutual inclination is consistent with the range of critical inclinations for this mechanism to operate, $39.2\degree < i_{\rm AB-b} < 140.8\degree$. The masses, eccentricities, and orbital periods also imply Kozai-Lidov oscillation periods less than the age of the system, $\log(\tau_{\rm KL}/{\rm Myr}) = 1.7^{+0.4}_{-0.5}$\,dex (Eq.~1; \citealp{2007ApJ...669.1298F}). Under the conservative assumption that \vhs~AB's initial eccentricity was zero, its maximum eccentricity attainable from Kozai-Lidov cycles is $[1-(5/3)\cos^2{i_{\rm AB-b, initial}}]^{-1/2}$ \citep{2007ApJ...669.1298F}. To achieve the observed $e_{\rm AB} = 0.883$ would thus have required an initial misalignment of $>$68.7\degree or $<$111.3\degree, the latter of which is in excellent agreement with our measured mutual inclination.

Whatever might have caused an initial misalignment between the two orbital planes may also be responsible for the unusual configuration of this system. The companion mass ratio relative to the inner binary is quite low ($M_{\rm b}/M_{\rm AB} = 0.08$--0.10), especially for a multiple system with total mass $<$0.2\,\Msun\ \citep[e.g.,][]{2007prpl.conf..427B}. At higher masses, the formation of such systems has been suggested to be due to the disintegration of high-order multiples at young ages \citep[e.g.,][]{2009MNRAS.392..413S,2015AJ....149..145R}, although such systems should be quite rare \citep[e.g.,][]{2012MNRAS.419.3115B}.

One final clue to the origins of the \vhs\ system is the eccentricity of the wide companion's orbit. Its periastron distance of $112^{+21}_{-25}$\,au is consistent with a more compact initial configuration for the system that led to \vhs~b being scattered onto a wide, eccentric, and misaligned orbit.

\section{Summary}

We measured high-precision relative astrometry of all three components of the \vhs\ system. Our orbital analysis yields a total dynamical mass of the inner binary ($M_{\rm A}+M_{\rm B} = 0.141\pm0.008$\,\Msun), high eccentricities for both the inner and outer orbits, and a mutual inclination of $116\pm16$\degree\ between them. We thus confirmed that the host binary may be a pair of brown dwarfs, derived their integrated-light luminosity from SED-fitting, and measured a cooling age of $140\pm20$\,Myr from \citet{2008ApJ...689.1327S} hybrid evolutionary models. We found that at such as young age, \vhs~b has a sufficiently low luminosity that it may be below the deuterium-fusing mass boundary or, only slightly more likely, that it is massive enough to have depleted its deuterium long ago. Regardless of the mass of \vhs~b, the orbital architecture implies a dynamical origin, perhaps from the disintegration of a high-order multiple or scattering within a protostellar disk.

If \vhs~b is indeed below the D-fusion mass boundary, then molecular absorption bands from D-bearing isotopologues of water (HDO) and methane (CH$_3$D) may be detectable in high-S/N 3--5~\micron\ {\sl JWST} spectra \citep[e.g.,][]{2019ApJ...882L..29M}. We also anticipate that similar observations will be possible for the other rare triple systems with planetary-mass companions for which substellar cooling ages are possible \citep[e.g., 2MASS~J0249$-$0557;][]{2018AJ....156...57D}.

%%%%%%%%%%%%%%%%%%%%%%%%%%%%%%%%%%%%%%%%%%%%%%%%%%
\section*{Acknowledgements}

We are grateful to the anonymous referee for prompt and thoughtful comments that improved our manuscript.
T.~Dupuy acknowledges support from UKRI STFC AGP grant ST/W001209/1.
This research was funded in part by the Gordon and Betty Moore Foundation through grant GBMF8550 to M.~Liu.
A.~Sanghi acknowledges support from the Research Experience for Undergraduate program at the Institute for Astronomy, University of Hawaii, Manoa funded through NSF grant \#2050710.
We thank Spencer Hurt for the BT-Settl models used in the bolometric luminosity calculation.
The data presented herein were obtained at the W.M.\ Keck Observatory, which is operated as a partnership between the California Institute of Technology, the University of California, and NASA. The Observatory was made possible by the generous financial support of the W.M.\ Keck Foundation. 
This work has made use of data from the European Space Agency (ESA) mission Gaia, processed by the Gaia Data Processing and Analysis Consortium (DPAC). Funding for the DPAC has been provided by national institutions, in particular the institutions participating in the Gaia Multilateral Agreement.
The authors wish to recognize and acknowledge the very significant cultural role and reverence that the summit of Maunakea has always had within the indigenous Hawaiian community. We are most fortunate to have the opportunity to conduct observations from this mountain. 
For the purpose of open access, the author has applied a Creative Commons Attribution (CC BY) licence to any Author Accepted Manuscript version arising from this submission.

%%%%%%%%%%%%%%%%%%%%%%%%%%%%%%%%%%%%%%%%%%%%%%%%%%
\section*{Data Availability}

All of our NIRC2 data are available on the Keck Observatory Archive (KOA), which is operated by the W.\ M.\ Keck Observatory and the NASA Exoplanet Science Institute (NExScI), under contract with the National Aeronautics and Space Administration.
We include configuration files for our orbit analysis in the supplemental data.

%%%%%%%%%%%%%%%%%%%% REFERENCES %%%%%%%%%%%%%%%%%%

%%%%%%%%%%%%%%%%%%%%%%%%%%%%%%%%%%%%%%%%%%%%%%%%%%

\label{lastpage}
\end{document}